\documentclass[12pt]{article}

\begin{document}

\begin{titlepage}

\rightline{hep-ph/0506093}

\vskip 2cm

\centerline{\large \bf {Axino dark matter in brane world cosmology}}

\vskip 1cm

\centerline{G. Panotopoulos}

\vskip 1cm

\centerline{Department of Physics, University of Crete}

\vskip 0.2 cm

\centerline{Heraklion, Crete, Hellas}

\vskip 0.2 cm

\centerline{email:{\it panotop@physics.uoc.gr}}

\begin{abstract}

We discuss dark matter in the brane world scenario. We work in the Randall-Sundrum type II brane world and assume that the lightest supersymmetric particle is the axino. We find that the axinos can play the role of cold dark matter in the universe, provided that the five-dimensional Planck mass is bounded both from below and from above. This is possible for higher reheating temperatures compared to the conventional four-dimensional cosmology due to a novel expansion law for the universe.

\end{abstract}

\end{titlepage}

\newpage

There are good theoretical reasons for which particle physics proposes that new exotic particles must exist. The most compelling solution to the strong CP-problem of quantum chromodynamics (QCD), which can be stated as ``why is the $\Theta$ parameter in QCD so small?'', is the one proposed by Peccei and Quinn \cite{quinn}. An additional global, chiral symmetry is introduced, now known as Peccei-Quinn (PQ) symmetry, which is spontaneously broken at the PQ scale $f_{\alpha} \geq 10^8 \: GeV$ \cite{kim}. Since $U(1)_{PQ}$ is a spontaneously broken global symmetry, there must be a Nambu-Goldstone boson associated with this symmetry. However, because $U(1)_{PQ}$ suffers from a chiral anomaly, this boson is not massless but acquires a small mass. The pseudo Nambu-Goldstone boson associated with this spontaneous symmetry breaking is the axion \cite{wilczek}, which has not yet been detected. On the other hand, supersymmetry (SUSY) is an ingredient that appears in many theories for physics beyond the standard model. SUSY solves the hierarchy problem and predicts that every particle we know should be escorted by its superpartner. The axino is the superpartner of the axion. In order for the supersymmetric solution of the hierarchy problem to work, it is necessary that the SUSY becomes manifest at relatively low energies, less than a few $TeV$, and therefore the required superpartners must have masses below this scale (for supersymmetry and supergravity see e.g. \cite{nilles}).

One of the theoretical problems in modern cosmology is to understand the nature of cold dark matter in the universe. There are good reasons, both observational and theoretical, to suspect that a fraction of $0.22$ of the energy density in the universe is in some unknown ``dark'' form \cite{spergel}. Many lines of reasoning suggest that the dark matter consists of some new, as yet undiscovered, weakly-interacting massive particle (WIMP), which experiences neither electromagnetic nor color interactions. In SUSY models which are realized with R-parity conservation the lightest supersymmetric particle (LSP) is stable. A popular candidate for cold dark matter is the LSP, provided that it is electrically and color neutral. Certainly the most theoretically developed WIMP is the lightest neutralino. However, there are other WIMP candidates as well, like for example the gravitino and the axino. In this work we assume that the axino is the LSP. Axinos are special because they have unique properties: They are very weekly interacting and their mass can span a wide range, from very small ($ \sim eV$) to large ($ \sim GeV$) values. What is worth stressing is that, in contrast to the neutralino and the gravitino, axino mass does not have to be of the order of the SUSY breaking scale in the visible sector, $M_{SUSY} \sim 100GeV-1TeV$. The first paper to show that the axinos can be CDM was Ref. \cite{ros}. There are however some early works on axino cosmology (see e.g. \cite{gabbiani} \cite{turner}).

We believe that some time in its early history, the universe experienced an inflationary phase \cite{kolb}. According to the inflationary paradigm, during the slow-roll phase of inflation the universe undergoes a rapid expansion, and consequently any initial population of axinos is diluted away. After slow-roll a reheating phase follows and leads the universe to the radiation era of the standard hot Big-Bang cosmology of temperature $T_R$. As the PQ symmetry is restored at $f_{\alpha}$, we consider only values of $T_R$ up to the PQ scale, which we take to be $f_{\alpha}=10^{11} \: GeV$. Another important scale is the temperature $T_D$ at which axinos decouple from the thermal bath. For $T_R > T_D$, there has been an early phase in which axinos were in thermal equilibrium with the thermal bath. The axino density parameter is then given by the equilibrium number density \cite{steffen}
\begin{equation}
\Omega_{\tilde{\alpha}} h^2 \sim \frac{m_{\tilde{\alpha}}}{2 \: keV}
\end{equation}
If we require that $\Omega_{\tilde{\alpha}} h^2 \sim 0.1$ then the axino mass $m_{\tilde{\alpha}} \sim 0.2 \; keV$. For an axino mass in the range $m_{\tilde{\alpha}} \leq 1 \: keV$, $1 \: keV \leq m_{\tilde{\alpha}} \leq 100 \: keV$ and $m_{\tilde{\alpha}} \geq 100 \: keV$, we refer to hot, warm and cold axino dark matter respectively. So we see that for $T_R > T_D$, axinos can only be hot dark matter.
For $T_R < T_D$, the axinos are out of thermal equilibrium so that the production mechanisms have to be considered in detail.

In order to generate large enough abundance of axinos, one needs to repopulate the universe (after inflation) with them. There are two generic ways of achieving this. First, they can be generated through thermal production (TP), namely via scattering and decay processes of ordinary particles and sparticles still in the thermal bath. Second, axinos may also be produced via non-thermal production mechanisms (NTP) possibly present during the reheating phase. In \cite{covi} the authors considered both NTP and TP and they found that $T_R$ had to be relatively low, below some $10^6 \: GeV$. However NTP mechanisms are strongly model dependent and we shall not consider them here. In \cite{steffen} the authors using specific techniques (the hard thermal loop resummation technique \cite{htl} together with the Braaten-Yuan prescription \cite{yuan}) computed the thermal production rate of axinos in supersymmetric QCD  and evaluated the relic axino abundance. They found that axinos provide the density of cold dark matter observed by WMAP for relatively small reheating temperature after inflation $T_R \leq 10^6 \: GeV$, essentially in agreement with \cite{covi}. Such a low reheating temperature excludes some models for inflation and the baryon asymmetry in the universe has to be explained by a mechanism that works efficiently at relatively small temperatures, excluding thermal leptogenesis \cite{yanagida}.  

The purpose of this letter is to show that this fact can be cured in the context of the brane world scenario. Our brane world model is the Randall-Sundrum type II model (RSII) \cite{randall}, and in fact its supersymmetric extended model \cite{pomarol}. However, the cosmological solution of this extended model is the same as that in the non-supersymmetric model, since the Einstein's equations belong to the bosonic part. The RSII model offers a novel expansion law for the observable four-dimensional universe. We find that the axino abundance today is proportional to the transition temperature, at which the modified expansion law in the brane world cosmology is connecting to the standard one, rather than the reheating temperature after inflation as in the standard cosmology. This means that even though the reheating temperature can be very high, the axinos can play the dominant part of the cold dark matter in the universe.  

Let us see in more detail the thermal production of axinos (in standard cosmology). We assume that after inflation axinos are far from thermal equilibrium. With the axino number density $n_{\tilde{\alpha}}$ being much smaller than the photon number density $n_{\gamma}$, the evolution of $n_{\tilde{\alpha}}$ with cosmic time $t$ can be described by the Boltzmann equation
\begin{equation}
\frac{dn_{\tilde{\alpha}}}{dt}+3 H n_{\tilde{\alpha}}=C_{\tilde{\alpha}}
\end{equation}
where $C_{\tilde{\alpha}}$ is the collision term, while the second term on the left-hand side accounts for the dilution of the axinos due to the expansion of the universe described by the Hubble parameter $H$.
It is convenient to define the dimensionless quantity
\begin{equation}
Y_{\tilde{\alpha}}=\frac{n_{\tilde{\alpha}}}{s}
\end{equation}
where $s$ is the entropy density for the relativistic degrees of freedom in the primordial plasma
\begin{equation}
s(T)=h_{eff}(T)  \frac{2 \pi^2}{45} T^3
\end{equation}
with $h_{eff}(T)  \simeq g_{eff}(T)$ in the radiation dominated epoch and $g_{eff}$ counts the total number of effectively massless degrees number of freedom (those species with mass $m_{i} \ll T$). When all the degrees of freedom are relativistic $g_{eff}=915/4=228.75$. Replacing the cosmic time $t$ with the temperature $T$, the number density $n_{\tilde{\alpha}}$ with the number-to-entropy ratio $Y_{\tilde{\alpha}}$ and using conservation of the entropy per comoving volume (see for example the first Ref. in \cite{kolb}), the Boltzmann equation can be cast into the form
\begin{equation}
\frac{dY_{\tilde{\alpha}}}{dT}=\frac{C_{\tilde{\alpha}}(T)}{T s(T) H(T)}
\end{equation}
where $H(T)$ is the Hubble parameter as a function of the temperature $T$ for the radiation dominated era
\begin{equation}
H(T)=\sqrt{\frac{\pi^2 g_{eff}}{90}} \frac{T^2}{M_{pl}}
\end{equation}
with $M_{p}=2.4 \times 10^{18} \: GeV$ is the reduced Planck mass. In terms of the number-to-entropy ratio $Y_{\tilde{\alpha}}$, the axino density parameter is given by
\begin{equation}
\Omega h^2=\frac{\rho_{\tilde{\alpha}} h^2}{\rho_{cr}}=\frac{m_{\tilde{\alpha}} n_{\tilde{\alpha}} h^2}{\rho_{cr}}=\frac{m_{\tilde{\alpha}} Y_{\tilde{\alpha}} s(T_0) h^2}{\rho_{cr}}
\end{equation}
Here we make use of the following values
\begin{eqnarray}
T_0 & = & 2.73 K=2.35 \times 10^{-13} \: GeV \\
h_{eff}(T_0)& = & 3.91 \\
\rho_{rc}/h^2& = & 8.1 \times 10^{-47} \: GeV^4
\end{eqnarray}
The collision term $C_{\tilde{\alpha}}$ has been computed in supersymmetric QCD by the authors of Ref. \cite{steffen}
\begin{equation}
C_{\tilde{\alpha}}(T)=\frac{3 \zeta(3) (N_{c}^2-1) g^6 T^6}{4096 \pi^7 f_{\alpha}^2} \: \left( 0.4336 n_{f}+(N_{c}+n_{f}) \: ln \left( \frac{1.38 T^2}{m_g}  \right)   \right)
\end{equation}
where $N_{c}=3$, $n_{f}=6$ and $g$ is the QCD coupling constant
\begin{equation}
g(T)=\left( \frac{1}{4 \times \pi \times 0.118}+\frac{3}{8 \pi^2} \: ln \left( \frac{T}{M_{Z}}  \right) \right)^{-1/2}
\end{equation}
The thermal axino production proceeds basically during the hot radiation dominated epoch, that is at temperatures above the one at matter-radiation equality $T_{eq}$. Integrating the Boltzmann equation the axino yield at the present temperature of the universe $T_0$ is given by
\begin{equation}
Y_{\tilde{\alpha}}(T_0)=\int _{T_{eq}} ^{T_R} dT \: \frac{C_{\tilde{\alpha}}}{T s(T) H(T)} \simeq \frac{C_{\tilde{\alpha}}(T_R)}{s(T_R) H(T_R)}
\end{equation}
and finally the axino density parameter is obtained
\begin{equation}
\Omega h^2=5.5 g^6 ln\left( \frac{1.108}{g} \right ) \left ( \frac{m_{\tilde{\alpha}}}{0.1 GeV} \right ) \left ( \frac{T_R}{10^4 GeV} \right ) \left( \frac{10^{11} \: GeV}{f_{\alpha}} \right )^2
\end{equation}
Considering $f_{\alpha}=10^{11} \: GeV$, axinos can be cold dark matter for masses $m_{\tilde{\alpha}} \geq 100 \: keV$ and reheating temperatures $T_R \leq 10^{6} \: GeV$ \cite{steffen}.

Recently the brane world models have been attracting a lot of attention as a novel higher dimensional theory. In these models, it is assumed that the standard model particles are confined on a 3-brane while gravity resides in the whole higher dimensional spacetime. The model first proposed by Randall and Sundrum (RSII) \cite{randall}, is a simple and interesting one, and its cosmological evolutions have been intensively investigated \cite{langlois}. According to that model, our 4-dimensional universe is realized on the 3-brane with a positive tension located at the UV boundary of 5-dimensional AdS spacetime. In the bulk there is just a cosmological constant $\Lambda_{5}$, whereas on the brane there is matter with energy-momentum tensor $\tau_{\mu \nu}$. Also, the five dimensional Planck mass is denoted by $M_{5}$ and the brane tension is denoted by $T$.
If Einstein's equations hold in the five dimensional bulk, then it has been shown in \cite{shir} that the effective four-dimensional Einstein's equations induced on the brane can be written as 
\begin{equation}
G_{\mu \nu}+\Lambda_{4} g_{\mu \nu}=\frac{8 \pi}{m_{pl}^2} \tau_{\mu \nu}+(\frac{1}{M_{5}^3})^2 \pi_{\mu \nu}-E_{\mu \nu}
\end{equation}
where $g_{\mu \nu}$ is the induced metric on the brane, $\pi_{\mu \nu}=\frac{1}{12} \: \tau \: \tau_{\mu \nu}+\frac{1}{8} \: g_{\mu \nu} \: \tau_{\alpha \beta} \: \tau^{\alpha \beta}-\frac{1}{4} \: \tau_{\mu \alpha} \: \tau_{\nu}^{\alpha}-\frac{1}{24} \: \tau^2 \: g_{\mu \nu}$, $\Lambda_{4}$ is the effective four-dimensional cosmological constant, $m_{pl}$ is the usual four-dimensional Planck mass and $E_{\mu \nu} \equiv C_{\beta \rho \sigma} ^\alpha \: n_{\alpha} \: n^{\rho} \: g_{\mu} ^{\beta} \: g_{\nu} ^{\sigma}$ is a projection of the five-dimensional Weyl tensor $C_{\alpha \beta \rho \sigma}$, where $n^{\alpha}$ is the unit vector normal to the brane.
The tensors $\pi_{\mu \nu}$ and $E_{\mu \nu}$ describe the influense of the bulk in brane dynamics. The five-dimensional quantities are related to the corresponding four-dimensional ones through the relations
\begin{equation}
m_{pl}=4 \: \sqrt{\frac{3 \pi}{T}} \: M_{5}^3
\end{equation}
and
\begin{equation}
\Lambda_{4}=\frac{1}{2 M_{5}^3} \left( \Lambda_{5}+\frac{T^2}{6 M_{5}^3} \right )
\end{equation}
In a cosmological model in which the induced metric on the brane $g_{\mu \nu}$ has the form of  a spatially flat Friedmann-Robertson-Walker model, with scale factor $a(t)$, the Friedmann-like equation on the brane has the generalized form (see e.g. the second Ref. in \cite{langlois})
\begin{equation}
H^2=\frac{\Lambda_{4}}{3}+\frac{8 \pi}{3 m_{pl}^2}  \rho+\frac{1}{36 M_{5}^6} \rho^2+\frac{C}{a^4}
\end{equation}
where $C$ is an integration constant arising from $E_{\mu \nu}$. The cosmological constant term and the term linear in $\rho$ are familiar from the four-dimensional conventional cosmology. The extra terms, i.e the ``dark radiation'' term and the term quadratic in $\rho$, are there because of the presense of the extra dimension. Adopting the Randall-Sundrum fine-tuning
\begin{equation}
\Lambda_{5}=-\frac{T^2}{6 M_{5}^3}
\end{equation}
the four-dimensional cosmological constant vanishes. So the generalized Friedmann equation takes the final form
\begin{equation}
H^2=\frac{8 \pi G}{3} \rho \left (1+\frac{\rho}{\rho_0} \right )+\frac{C}{a^4}
\end{equation}
where
\begin{equation}
\rho_0=96 \pi G M_{5}^6
\end{equation}
with $G$ the Newton's constant. The second term proportional to $\rho^2$ and the dark radiation are new ingredients in the brane world cosmology and lead to a non-standard expansion law. The dark radiation term is severly constrained by the success of the Big Bang Nucleosynthesis (BBN), since the term behaves like an additional radiation at the BBN era \cite{orito}. So, for simplicity, we neglect the term in the following analysis. The second term is also constrained by the BBN, which is roughly estimated as $M_{5} \geq 10 \: TeV$ \cite{cline}.

One can see that the evolution of the early universe can be divided into two eras. In the low energy regime $\rho \ll \rho_0$ the first term dominates and we recover the usual Friedmann equation of the conventional four-dimensional cosmology. In the high density regime $\rho_0 \ll \rho$ the second term dominates and we get an unconventional expansion law for the universe. In between there is a transition temperature $T_t$ in which $\rho(T_t)=\rho_0$. Using the transition temperature the generalized Friedmann-like equation (for the radiation era) can be rewritten in the form
\begin{equation}
H=H_{st} \: \sqrt{1+\frac{T^4}{T_t^4}}
\end{equation}
with $H_{st}$ the Hubble parameter in standard  4-dimensional Big-Bang cosmology. Assuming a transition temperature $T_R \gg T_t$ and $T_t \gg T_{eq}$, the following integral can be computed to a very good approximation
\begin{equation}
\int _{T_{eq}} ^{T_R} dT \: \frac{1}{\sqrt{1+\frac{T^4}{T_t^4}}}=\int _{T_{eq}} ^{T_t} dT \: \frac{1}{\sqrt{1+\frac{T^4}{T_t^4}}}+\int _{T_t} ^{T_R} dT \: \frac{1}{\sqrt{1+\frac{T^4}{T_t^4}}} \simeq 2 T_t
\end{equation}
Therefore, the axino yield resulting by integrating the Boltzmann equation in brane cosmology is
\begin{equation}
Y_{\tilde{\alpha}}(T_0)=\int _{T_{eq}} ^{T_R} dT \: \frac{C_{\tilde{\alpha}}(T)}{T s(T) H_{st}(T)} \: \frac{1}{\sqrt{1+\frac{T^4}{T_t^4}}} \simeq \frac{C_{\tilde{\alpha}}(T_R)}{T_R s(T_R) H_{st}(T_R)} \: 2 T_t
\end{equation}
So we see that essentially the reheating temperature $T_R$ in the axino parameter density $\Omega_{\tilde{\alpha}}$ is replaced by the transition temperature $T_t$. Therefore, axinos can play the role of cold dark matter in the universe for
\begin{equation}
m_{\tilde{\alpha}} \geq 100 \: keV, \; \; T_t \leq 2 \times 10^6 \: GeV
\end{equation}
independently of the reheating temperature. This is the main result in this letter. In this point we would like to stress the fact that the axino abundance does depend on the reheating temperature, but only through the coupling constant $g(T_{R})$. The function $g(T)$ is a very slow-varying function of the temperature (for example, $g(T=10^6 \: GeV)=0.986$ and $g(T=10^{10} \: GeV)=0.852$), so practically we can consider reheating temperatures of the order of $\sim 10^{10} \: GeV$ and neglect the dependence on it. Note that according to the analysis of \cite{okada}, with a transition temperature $T_t \leq 10^6 \: GeV$ the gravitino problem can be avoided. It is interesting to note that in order that the axinos can play the role of cold dark matter in the universe, the five-dimensional Planck mass $M_5$ can only take values in a range between an upper limit and a lower limit. If $M_5$ becomes too high, the transition temperature $T_t$ will be higher than $\sim 10^6 \: GeV$ and this sets an upper bound for $M_5$, $M_{5} \leq 2.9 \times 10^{10} \: GeV$. On the other hand, if $M_5$ becomes too low, the mass of the axinos will be larger than the mass of the next lighest supersymmetric particle. Assuming that the axino mass can be at most $m_{\tilde{\alpha}} \sim 100 \: GeV$, we get a lower bound for $M_5$, $M_{5} \geq 2.4 \times 10^6 \: GeV$. So we find a window for the five-dimensional Planck mass
\begin{equation}
2.4 \times 10^6 \: GeV \leq M_{5} \leq 2.9 \times 10^{10} \: GeV
\end{equation}

To illustrate the above ideas let us present a specific example. We consider the case in which the 5-dimensional Planck mass is $M_5=10^{10} \: GeV$. Then the transition temperature $T_t$ is found to be $T_t=4 \times 10^5 \: GeV$. We also assume that the reheating temperature is $T_R=10^{10} \: GeV$. If the axinos are to be the cold dark matter in the universe, their parameter density has to be $\Omega_{\tilde{\alpha}} h^2=0.113$. This happens for $m_{\tilde{\alpha}} \simeq 511 \: keV$. Our treatment is valid as long as axinos are never in thermal equilibtium after inflation. One can easily check that this is the case. For that we need to compare the Hubble parameter $H(T)$ to the rate $\Gamma(T)$ of the reaction that maintain the axinos in thermal equilibrium. For $T < f_{\alpha}$ the reaction rate is \cite{turner}
\begin{equation}
\Gamma \sim \frac{\alpha_{s}^3}{16 \pi f_{\alpha}^2} \: T^3
\end{equation}
We can see that after inflation $H(T) \gg \Gamma(T)$ for all the values of $M_{5}$ in the allowed range.

Summarizing, we have discussed dark matter in Randall-Sundrum type II brane world assuming that the axino is the LSP. We have seen that axinos can be the dominant part of the cold dark matter in the universe if their mass $m_{\tilde{\alpha}} \geq 100 \; keV$ and the transition temperature $T_t \leq 2 \times 10^{6} \: GeV$ independently of the reheating temperature $T_R$ after inflation (provided that $T_R \gg T_t$, which is true for $T_{R} \sim 10^{10} \: GeV$). Therefore, in contrast to the conventional 4-dimensional cosmology, high values for $T_R$ such as $10^{10} \: GeV$ are allowed, in accord to most inflationary models and baryogenesis through leptogenesis.


\textbf{Acknowlegements:} The author is greatful to T.N.Tomaras for useful discussions. Work supported by the Greek Ministry of education research program "Heraklitos" and by the EU grant MRTN-CT-2004-512194.

\end{document}